\documentclass[11pt,a4paper]{amsart}
\usepackage{geometry}                
\usepackage{graphicx}
\usepackage{amssymb}
\usepackage{epstopdf}
\usepackage{fullpage}
\usepackage{color}
\usepackage{float}
\usepackage{caption}
\usepackage[table]{xcolor}
\usepackage{tablefootnote}

\usepackage{url}
\usepackage{morefloats}

\usepackage{graphicx,amsmath,multicol,txfonts}

\newcommand{\bfmu} {\boldsymbol{\mu}}

\newcommand{\bfphi} {\boldsymbol{\phi}}

\newcommand{\bfTheta} {\boldsymbol{\Theta}}
\newcommand{\bfSigma} {\boldsymbol{\Sigma}}

\newcommand{\bfOmega} {\boldsymbol{\Omega}}

\newcommand{\bfM} {\mathbf{M}}

\newcommand{\bfS} {\mathbf{S}}

\newcommand{\bfY} {\mathbf{Y}}
\newcommand{\bfX} {\mathbf{X}}
\newcommand{\bfV} {\mathbf{V}}
\newcommand{\bfE} {\mathbf{E}}
\newcommand{\bfW} {\mathbf{W}}
\newcommand{\bfD} {\mathbf{D}}

\newcommand{\bfK} {\mathbf{K}}

\newcommand{\bfx} {\mathbf{x}}

\newcommand{\bfI} {\mathbf{I}}

\newcommand{\pr}{\mathsf{p}}

\newcommand{\Cov}{\mathsf{Cov}}

\newcommand{\reals}{\mathbb{R}}

\DeclareMathOperator{\diag}{diag}

\newcommand{\ground}{\mathcal{G}}

\newcommand{\nbd}{\mathsf{nbd}}
\newcommand{\bd}{\mathsf{bd}}
\newcommand{\normal}{\mathsf{N}}
\newcommand{\anormal}{\mathsf{AN}}

\newcommand{\Wis}{\mathsf{Wis}}

\newcommand{\Poi}{\mathsf{Poi}}

\newcommand{\Uni}{\mathsf{Uni}}
\newcommand{\GWis}{\mathsf{Wis}}

\newcommand{\cone}{\mathsf{P}}

\newtheorem{theorem}{Theorem}

\begin{document}

\title{Graphical Modeling of Spatial Health Data}

\author{Adrian Dobra}
\address{Department of Statistics, Department of Biobehavioral Nursing and Health Systems, Center for Statistics and the Social Sciences and Center for Studies in Demography and Ecology, University of Washington, Box 354322, Seattle, WA 98195}
\email{adobra@uw.edu}

\begin{abstract}
The literature on Gaussian graphical models (GGMs) contains two equally rich and equally significant domains of research efforts and interests.  The first research domain relates to the problem of graph determination. That is, the underlying graph is unknown and needs to be inferred from the data. The second research domain dominates the applications in spatial epidemiology. In this context GGMs are typically referred to as Gaussian Markov random fields (GMRFs). Here the underlying graph is assumed to be known: the vertices correspond to geographical areas, while the edges are associated with areas that are considered to be neighbors of each other (e.g., if they share a border).  We introduce multi-way Gaussian graphical models that unify the statistical approaches to inference for spatiotemporal epidemiology with the literature on general GGMs. The novelty of the proposed work consists of the addition of the G-Wishart distribution to the substantial collection of statistical tools used to model multivariate areal data.  As opposed to fixed graphs that describe geography, there is an inherent uncertainty related to graph determination across the other dimensions of the data. Our new class of methods for spatial epidemiology allow the simultaneous use of GGMs to represent known spatial dependencies and to determine unknown dependencies in the other dimensions of the data.
\end{abstract}

\maketitle

\tableofcontents

\section{Introduction}

Graphical models \cite{whittaker1990,lauritzen_1996} that encode multivariate independence and conditional independence relationships among observed variables $X=(X_{1},\ldots,X_{p})$ have a widespread use in major scientific areas (e.g., biomedical  and social sciences). In particular, a Gaussian graphical model (GGM) is obtained by setting off-diagonal elements of the precision matrix $K = \Sigma^{-1}$ to zero of a $p$-dimensional multivariate normal model \cite{dempster_1972}. Employing a GGM instead of a multivariate normal model leads to a significant reduction in the number of parameters that need to be estimated if most elements of $K$ are constrained to be zero and $p$ is large. A pattern of zero constraints in $K$ can be recorded as an undirected graph $G=(V,E)$ where the set of vertices $V=\{1,2,\ldots,p\}$ represent observed variables, while the set of edges $E\subset V\times V$ link all the pairs of vertices that correspond to off-diagonal elements of $K$ that have not been set to zero. The absence of an edge between $X_{v_1}$ and $X_{v_2}$ corresponds with the conditional independence of these two
random variables given the rest and is denoted by $X_{v_1}\Perp X_{v_2}\mid X_{V\setminus \{ v_1,v_2\}}$ \cite{wermuth_1976}. This is called the pairwise Markov property relative to $G$, which in turn implies the local as well as the global Markov properties relative to $G$ \cite{lauritzen_1996}. The local Markov property plays a key role since it gives the regression model induced by $G$ on each variable $X_v$. More explicitly, consider the neighbors of $v$ in $G$, that is, the set of vertices $v'\in V$ such that $(v,v')\in E$. We denote this set by $\mbox{bd}_G(v)$. The local Markov property relative to G says that $X_{v}\Perp X_{V\setminus \{ \{v\}\cup \mbox{bd}_G(v)\} }\mid X_{\mbox{bd}_G(v)}$. This statement is precisely the statement we make when we drop the variables $\{X_{v'}:v'\in V\setminus \mbox{bd}_G(v)\}$ from the regression of $X_v$ on $\{ X_{v'}:v'\in V\setminus \{v\}\}$.\\
\indent The literature on GGMs contains two equally rich and significant domains of research. The first research domain relates to the problem of graph determination. That is, the underlying graph is unknown and needs to be inferred from the data. Frequentist methods estimate $K$ and $\Sigma$ given one graph that is best supported by the data in the presence of sparsity constraints that penalize for increased model complexity (i.e., for the addition of extra edges in the graph). Among numerous notable contributions we mention the regularization methods of \cite{meinshausen_buhlman_2006,yuan_lin_2007,bickel_levina_2008,friedman-et-2008} as well as the simultaneous confidence intervals of \cite{drton_perlman_2004}. Bayesian methods proceed by imposing suitable prior distributions for $K$ or $\Sigma$ \cite{leonard_hsu_1992,yang:berger:1994,daniels:kass:1999,barnard_et_2000,smith_kohn_2002,liechty_et_2004,rajaratnam_et_2008}. Inference can be performed based on the best model, i.e. the graph having the highest posterior probability, or by Bayesian model averaging \cite{kass_raftery_1995} over all $2^{p(p-1)/2}$ possible graphs using Markov chain Monte Carlo (MCMC) approaches \cite{giudici_green_1999,dellaportas-et-2003,wong-et-2003}. As the number of graphs grows, MCMC methods are likely to visit only subsets of graphs that have high posterior probabilities. To this end, various papers \cite{jones_et_2005,scott_carvalho_2008,lenkoski-dobra-2010} have proposed stochastic search methods for fast identification of these high posterior probability graphs.\\
\indent The second research domain on GGMs dominates the applications in spatial epidemiology. In this context GGMs are referred to as Gaussian Markov random fields (GMRFs) \cite{Be74,Be75,BeKa95}. The underlying graph $G$ is assumed to be known: the vertices correspond to geographical areas, while the edges are associated with areas that are considered to be neighbors of each other (e.g., if they share a border). A GMRF is specified through the conditional distributions of each variable given the rest
\begin{eqnarray}\label{eq:fullcond}
 \left\{ p(X_{v}\mid X_{V\setminus \{v\}}):v\in V\right\},
\end{eqnarray}
which are assumed to be normal. The local Markov property leads to a further reduction in the set of full conditionals:
\begin{eqnarray} \label{eq:markovcond}
 p(X_{v}\mid X_{V\setminus \{v\}}) = p(X_{v}\mid X_{\mbox{bd}_{G}(v)}).
\end{eqnarray}
Since it is typically assumed that phenomena (e.g., the occurrence of a disease) taking place in one area influence corresponding phenomena taking place in the remaining areas only through neighbor areas, the set of reduced conditionals (\ref{eq:markovcond}) are employed to describe a full joint distribution of random spatial effects. GMRFs are conditional autoregressions (CAR) models that have a subclass called simultaneous autoregressions (SAR). For a comprehensive account of inference in CAR/SAR/GMRFs see \cite{Cr73,RuHe05,gelfand-et-2010}. Key questions relate to conditions in which a joint distribution determined by (\ref{eq:fullcond}) actually exists and, if it does, whether it is multivariate normal. This leads to particular parametric specifications for the set of conditionals (\ref{eq:fullcond}) and (\ref{eq:markovcond}) that are more restrictive than the general parametric specification of a GGM.\\
\indent In this chapter we examine the theoretical differences between GGMs and GMRFs. \cite{DoLeRo11} developed efficient MCMC methods for inference in univariate and matrix-variate GGMs, and subsequently employed these methods to construct Bayesian hierarchical spatial models for mapping multiple diseases. We extend their results to multi-way GGMs that can capture temporal dependencies in addition to several other relevant dimensions. We exemplify the construction of a Bayesian hierarchical spatiotemporal model based on three-way GGMs, and also present a related theoretical extension of multi-way GGMs to dynamic multi-way GGMs for array-variate time series.

\section{GGMs vs. GMRFs} \label{sec:contrast}

We consider a GGM defined by a graph $G=(V,E)$ for the multivariate normal distribution $\normal_{p}(0,K^{-1})$ of a vector $X=(X_{1},\ldots,X_{p})$. The precision matrix $K$ is constrained to belong to the cone $\cone_{G}$ of positive definite matrices such that $K_{ij}=0$ for all $(i,j)\notin E$. The full conditionals (\ref{eq:fullcond}) associated with each $X_{v}$ are expressed as a function of the elements of $K$ as follows:
\begin{eqnarray}\label{eq:normalcond}
 X_{v}\mid X_{V\setminus \{v\}}=x_{V\setminus \{v\}} \sim \normal(- \sum_{v^{\prime}\in \mbox{bd}_{G}(v)} (K_{vv^{\prime}}/K_{vv}) x_{v^{\prime}},1/K_{vv}).
\end{eqnarray}
Remark that the variables $X_{v^{\prime}}$ that are not linked by an edge with $X_{v}$ are dropped from the full conditional (\ref{eq:normalcond}) because $K_{vv^{\prime}}=0$. A GMRF with graph $G$ is parametrized through the full conditionals
\begin{eqnarray} \label{eq:fullcondraw}
 X_{v}\mid X_{V\setminus \{v\}}=x_{V\setminus \{v\}} \sim \normal(  \sum _{v^{\prime}\in \mbox{bd}_{G}(v)} \beta_{vv^{\prime}}x_{v^{\prime}},\sigma_{v}^{2}), \mbox{ for } v\in V.
\end{eqnarray}
The symmetry condition $\beta_{vv^{\prime}}\sigma^{2}_{v^{\prime}}=\beta_{v^{\prime}v}\sigma^{2}_{v}$ for all $v\ne v^{\prime}$ is necessary for the conditionals (\ref{eq:fullcondraw}) to define a proper GGM \cite{lauritzen_1996}. Additional constraints under which the set of regression parameters $\{(\beta_{vv^{\prime}},\sigma_{v}^{2})\}$ induce a proper precision matrix $K\in \cone_{G}$ with $K_{vv}=\sigma_{v}^{-2}$ and $K_{vv^{\prime}}=-\beta_{vv^{\prime}}\sigma_{v}^{-2}$ are given in \cite{BeKa95}. \cite{besag-york-mollie-1991} make use of a symmetric proximity matrix $W$ with $w_{vv}=0$, $w_{vv^{\prime}}>0$ if $v^{\prime}\in \mbox{bd}_{G}(v)$ and $w_{vv^{\prime}}=0$ if $v^{\prime}\notin \mbox{bd}_{G}(v)$. They define $\beta_{vv^{\prime}}=\rho w_{vv^{\prime}}/w_{v+}$ and $\sigma_{v}^{2}=\sigma^{2}/w_{v+}$, where $w_{v+}=\sum_{v^{\prime}}w_{vv^{\prime}}$. Here $\rho$ is referred to as a spatial autocorrelation parameter. With this choice, for each $\rho\in (-1,1)$ and $\sigma^{2}>0$, the GMRF specified by the full conditionals (\ref{eq:fullcondraw}) has a precision $K=\sigma^{2}(E_{W}-\rho W)^{-1}\in \cone_{G}$, where $E_{W}=\mbox{diag}\{w_{1+},\ldots,w_{p+}\}$. As such, this widely used parametrization of GMRFs is quite restrictive since not any matrix in the cone $\cone_{G}$ can be represented through the two parameters $\rho$ and $\sigma^{2}$ given a particular choice of $W$. This difficulty originates from the parametrization (\ref{eq:fullcondraw}) of a GGM. Instead, by imposing proper prior distributions for the precision matrix $K$, we avoid the unnecessary representation of a GGM as the set of full conditionals (\ref{eq:fullcondraw}). In particular, we use of the G-Wishart prior $\GWis_G(\delta,D)$ with density 
\begin{eqnarray} \label{eq:wishart}
p\left(K\mid G,\delta,D\right) & = &\frac{1}{I_G(\delta,D)}(\mbox{det}\; K)^{(\delta-2)/2}\exp\{-\frac{1}{2}\langle K,D\rangle\},
\end{eqnarray}
with respect to the Lebesgue measure on $\cone_G$
\cite{roverato_2002,atay-kayis_massam_2005,letac_massam_2007}. Here $\langle A,B\rangle = \mbox{tr}(A^{T}B)$ denotes the trace inner product. The normalizing constant $I_G(\delta,D)$ is finite provided $\delta>2$ and $D$ positive definite \cite{diaconis_ylvisaker_1979}. The G-Wishart prior $\GWis_G(\delta,D)$ is conjugate to the normal likelihood. For a thorough account of its numerical properties see \cite{lenkoski-dobra-2010} and the references therein.\\
\indent For applications in hierarchical spatial models, \cite{DoLeRo11} set $D=(\delta-2)\sigma^{2}(E_{W}-\rho W)^{-1}$ because, with this choice, the prior mode for $K$ is precisely $\sigma^{2}(E_{W}-\rho W)^{-1}$ \--- the precision matrix of a GMRF. The prior specification for the precision matrix can therefore be completed in a manner similar to the current work from the existent literature on GMRFs. We note that the G-Wishart prior for $K$ induces compatible prior distributions for the regression parameters (\ref{eq:fullcondraw}) \--- see \cite{DoHaJoNeYaWe04}. The advantage of  this representation of GGMs is a more flexible framework for GMRFs that allows their regression coefficients to be determined from the data rather than being fixed or allowed to vary as a function of only two parameters. 

\section{Multi-way Gaussian Graphical Models} \label{sec:multiway}

We develop a framework for analyzing datasets that are associated with a random $L$-dimensional array $X$. Such datasets are quite common in social and biomedical sciences. In particular, spatial epidemiology involves datasets recording SIRs of several diseases observed under different conditions at multiple time points. The notations, definitions and operators related to tensors that appear throughout are introduced in \cite{lathauwer-2000,kolda-2006}. The elements of the observed multi-way array are indexed by $\{ (i_{1},i_{2},\ldots,i_{L}):1\le i_{l}\le m_{l}\}$. The total number of elements of $\bfX$ is $m=\prod_{l=1}^{L}m_{l}$. We assume that $\bfX$ follows an array normal distribution 
 $$\mbox{vec}(\bfX)\mid \bfK \sim \normal_{m}(0,\bfK^{-1}),$$ 
\noindent whose $m\times m$ precision matrix $\bfK$ is separable across each dimension, i.e.
\begin{eqnarray}\label{eq:sepnorm}
 \bfK & = \bfK_{L}\otimes \bfK_{L-1}\otimes\ldots\otimes \bfK_{1}.
\end{eqnarray}
\noindent The $m_{l}\times m_{l}$ precision matrix $\bfK_{l}$ is associated with dimension $l$, while $\mbox{vec}(\bfX)$ is the vectorized version of $\bfX$. The separability assumption might seem restrictive in the sense that it captures only dependencies across each dimension of the data without directly taking into account the interactions that might exist among two, three or more dimensions. However, this assumption reduces the number of parameters of the distribution of $\mbox{vec}(\bfX)$ from $2^{-1}m(m+1)$ to $2^{-L}\prod_{l=1}^{L}m_{l}(m_{l}+1)$ which constitutes a substantial advantage when a sample size is small. The probability density of $\bfX$ as an array is (see \cite{hoff-2010})
\begin{eqnarray}\label{eq:arraynormal}
 \pr(\bfX\mid \bfK_{1},\ldots,\bfK_{L}) & = & (2\pi)^{-\frac{m}{2}}\left[\prod\limits_{l=1}^{L} (\mbox{det}\; \bfK_{l})^{\frac{1}{m_{l}}}\right]^{\frac{m}{2}} \exp\left\{ -\frac{1}{2}\langle \bfX,\bfX \times \{ \bfK_{1},\ldots,\bfK_{L}\}\rangle\right\},
\end{eqnarray}
where
$$
 \bfX \times \{ \bfK_{1},\ldots,\bfK_{L}\} = \bfX \times_{1} \bfK_{1} \times_{2}\ldots \times_{L} \bfK_{L},
$$
is the Tucker product. Here $\bfX \times_{l} \bfK_{l}$ is the $l$-mode product of the tensor $\bfX$ and matrix $\bfK_{l}$. We refer to (\ref{eq:arraynormal}) as the mean-zero $L$-dimensional array normal distribution $\anormal_{L}(\mathbf{0};\{m_{1},\bfK_{1}\},\ldots,\{m_{L},\bfK_{L}\})$.\\
\indent Most of the existent literature has focused on two-dimensional (or matrix-variate) arrays \--- see, for example, \cite{allen-tibshirani-2010,olshen-rajaratnam-2010} and the references therein. \cite{galecki-1994} studies the separable normal model (\ref{eq:sepnorm}) for $L=3$, while \cite{dawid-1981} presents theoretical results for matrix-variate distributions that includes (\ref{eq:sepnorm}) with $L=2$ as a particular case. \cite{hoff-2010} has proposed a Bayesian inference framework for model (\ref{eq:sepnorm}) for an arbitrary number $L$ of dimensions by assigning independent inverse-Wishart priors for the covariance matrices $\bfSigma_{l}=\bfK_{l}^{-1}$ associated with each dimension. Despite its flexibility and generality, the framework of \cite{hoff-2010} does not allow any further reduction in the number of parameters of model (\ref{eq:sepnorm}). To this end, we propose a framework in which each precision matrix $\bfK_{l}$ is constrained to belong to a cone $P_{G_{l}}$ associated with a GGM with graph $G_{l}\in \mathcal{G}_{m_{l}}$. We denote by $\mathcal{G}_{m_{l}}$ the set of undirected graphs with $m_{l}$ vertices. Sparse graphs associated with each dimension lead to sparse precision matrices, hence the number of parameters that need to be estimated could be significantly smaller than $2^{-L}\prod_{l=1}^{L}m_{l}(m_{l}+1)$. A similar framework has been proposed in \cite{wang_west_2009} for matrix-variate data ($L=2$) and for row and column graphs restricted to the class of decomposable graphs. Our framework is applicable for any number of dimensions and allows arbitrary graphs (decomposable and non-decomposable) to be associated with each precision matrix $\bfK_{l}$.\\
\indent The prior specification for $\{\bfK_{l}\}_{l=1}^{L}$ must take into account the fact that two precision matrices are not uniquely identified from their Kronecker product which means that, for any $z>0$ and $l_{1}\ne l_{2}$,
$$
\bfK_{L}\otimes \cdots \otimes\left(z^{-1}\bfK_{l_{1}}\right)\otimes\cdots \otimes\left(z \bfK_{l_{2}}\right) \otimes\cdots\otimes \bfK_{1}= \bfK_{L}\otimes \cdots \otimes \bfK_{l_{1}} \otimes\cdots \otimes \bfK_{l_{2}} \otimes\cdots\otimes \bfK_{1}
$$
\noindent represents the same precision matrix for $\mbox{vec}(\bfX)$. We follow the basic idea laid out in \cite{wang_west_2009} and impose the constraints
\begin{eqnarray} \label{eq:k11}
 (\bfK_{l})_{11}=1, \mbox{ for } l=2,\ldots,L.
\end{eqnarray}
Furthermore, we define a prior for $\bfK_{l}$, $l\ge 2$, through parameter expansion by assuming a G-Wishart prior $\GWis_{G_{l}}(\delta_{l},\bfD_{l})$ for the matrix $z_{l} \bfK_{l}$ with $z_{l}>0$, $\delta_{l}>2$ and $\bfD_{l}\in \cone_{G_{l}}$.\noindent We denote $G_{l}=(V_{l},E_{l})$, where $V_{l}=\{ 1,2,\ldots,m_{l}\}$ are vertices and $E_{l}\subset V_{l}\times V_{l}$ are edges. We consider the Cholesky decompositions of the precision matrices from (\ref{eq:sepnorm}),
\begin{eqnarray} \label{eq:choldecomp}
 \bfK_{l} = \bfphi_{l}^{T}\bfphi_{l},
\end{eqnarray}
where $\bfphi_{l}$ is an upper triangular matrix with $(\bfphi_{l})_{ii}>0$, $1\le i\le m_{l}$.  \cite{roverato_2002} proves that the set $\nu(G_{l})$ of the free elements of $\bfphi_{l}$ consists of the diagonal elements together with the elements that correspond with the edges of $G_{l}$, i.e.
$$
\nu \left( G_{l}\right) = \{(i,i):i\in V_{l}\} \cup \{ (i,j):i<j \mbox{ and } (i,j)\in E_{l}\}.
$$
Once the free elements of $\bfphi_{l}$ are known, the remaining elements are also known. Specifically, $(\bfphi_{l})_{1j}=0$ if $j\ge 2$ and $(1,j)\notin E_{l}$. We also have
$$
 (\bfphi_{l})_{ij} = -\frac{1}{(\bfphi_{l})_{ii}}\sum\limits_{k=1}^{i-1}(\bfphi_{l})_{ki}(\bfphi_{l})_{kj},
$$
for $2\le i<j$ and $(i,j)\notin E_{l}$. The determination of the elements of $\bfphi_{l}$ that are not free based on the elements of $\bfphi_{l}$ that are free is called the completion of $\bfphi_{l}$ with respect to $G_{l}$ \cite{roverato_2002,atay-kayis_massam_2005}. It is useful to remark that the free elements of $\bfphi_{l}$ fully determine the matrix $K_{l}$. The development of our framework involves the Jacobian of the transformation that maps $\bfK_{l}\in \cone_{G_{l}}$ to the free elements of $\bfphi_{l}$ \cite{roverato_2002}:
$$
 J(\bfK_{l} \rightarrow \bfphi_{l}) = 2^{m_{l}}\prod_{i=1}^{m_{l}}(\bfphi_{l})_{ii}^{d^{G_{l}}_{i}+1},
$$
where $d^{G_{l}}_{i}$ is the number of elements in $\bd_{G_{l}}(i)\cap \{i+1,\ldots,m_{l}\}$ and $\bd_{G_{l}}(i)=\{ j:(i,j)\in E_{l}\}$ is the boundary of vertex $i$ in $G_{l}$.\\
\indent Our proposed prior specification for the separable normal model (\ref{eq:arraynormal}) is
\begin{eqnarray} \label{eq:sepnormprior}
 \bfK_{1}\mid \delta_{1},\bfD_{1} \sim \Wis_{G_{1}}(\delta_{1},\bfD_{1}),\; (z_{l}\bfK_{l})\mid  \delta_{l},\bfD_{l} & \sim & \Wis_{G_{l}}(\delta_{l},\bfD_{l}), \mbox{ for } l=2,\ldots,L.
\end{eqnarray}
The prior for $\bfK_{1}$ is
\begin{eqnarray} \label{eq:gwishart}
 \pr\left( \bfK_{1}\mid G_{1}\right) = \frac{1}{I_{G_{1}}\left(\delta_{1},\bfD_{1}\right)}\left(\mbox{det}\; \bfK_{1}\right)^{\frac{\delta_{1}-2}{2}}\exp\left\{-\frac{1}{2}\langle \bfK_{1},\bfD_{1}\rangle\right\},
\end{eqnarray}
while the joint prior for $\left( z_{l}, \bfK_{l}\right)$ is
\begin{eqnarray}\label{eq:zkprior}
 \pr\left( z_{l}, \bfK_{l}\mid G_{l}\right) = \frac{1}{I_{G_{l}}\left(\delta_{l},\bfD_{l}\right)}\left(\mbox{det}\; \bfK_{l}\right)^{\frac{\delta_{l}-2}{2}}\exp\left\{-\frac{1}{2}\langle \bfK_{l},z_{l} \bfD_{l}\rangle\right\} z_{l}^{\frac{m_{l}(\delta_{l}-2)}{2}+|\nu\left( G_{l}\right)|-1},
\end{eqnarray}
\noindent for $l=2,\ldots,L$.

\section{Inference in Multi-way GGMs} \label{sec:mcmc}

We assume that the observed samples $\mathcal{D}=\{\bfx^{(1)},\ldots,\bfx^{(n)}\}$ are independently generated from the mean-zero array normal distribution $\anormal_{L}(\mathbf{0};\{m_{1},\bfK_{1}\},\ldots,\{m_{L},\bfK_{L}\})$. The resulting likelihood is expressed by introducing an additional dimension $m_{L+1}=n$ with precision matrix $\bfK_{L+1}=\bfI_{n}$, where $\bfI_{n}$ is the $n\times n$ identity matrix. We see $\mathcal{D}$ as a $m_{1}\times\ldots\times m_{L+1}$ array that follows an array normal distribution $\anormal_{L+1}(\mathbf{0};\{m_{1},\bfK_{1}\},\ldots,\{m_{L+1},\bfK_{L+1}\})$. Furthermore, we define $\bfphi_{L+1}=\bfI_{n}$. The Cholesky decompositions (\ref{eq:choldecomp}) of the precision matrices $\bfK_{l}$ give the following form of the likelihood of $\mathcal{D}$:
\begin{eqnarray}\label{eq:lik}
 \pr(\mathcal{D}\mid \bfK_{1},\ldots,\bfK_{L}) & \propto & \left[ \prod_{l=1}^{L} (\mbox{det}\; \bfK_{l})^{\frac{1}{m_{l}}}\right]^{\frac{mn}{2}} \exp\left\{ -\frac{1}{2} \| \mathcal{D} \times \{ \bfphi_{1},\ldots,\bfphi_{L+1}\}\|\right\},
\end{eqnarray}
\noindent where $\| \bfY\| = \langle \bfY,\bfY\rangle$ is the array norm \cite{kolda-2006}. Simple calculations show that the part of the likelihood (\ref{eq:lik}) that depends on the precision matrix $\bfK_{l}$ is written as:
\begin{eqnarray}\label{eq:likdiml}
\pr(\mathcal{D}\mid \bfK_{1},\ldots,\bfK_{L}) & \propto & (\mbox{det}\; \bfK_{l})^{\frac{mn}{2m_{l}}} \exp\left\{ -\frac{1}{2} \langle \bfK_{l},\mathbf{S}_{l}\rangle\right\},
\end{eqnarray} 
\noindent where $\mathbf{S}_{l}= \mathcal{D}^{[l]}_{(l)}\left(\mathcal{D}^{[l]}_{(l)}\right)^{T}$ and
$$
\mathcal{D}^{[l]} = \mathcal{D}\times_{1} \bfphi_{1}\times_{2}\ldots\times_{l-1}\bfphi_{l-1}\times_{l}\bfI_{l}\times_{l+1}\bfphi_{l+1}\times_{l+2}\ldots\times_{L+1}\bfphi_{L+1}.
$$ 
Here $\bfY_{(l)}$ is the $l$-mode matricization of an array $\bfY$ \cite{kolda-2006}. We develop a Markov chain Monte Carlo sampler from the posterior distribution of precision matrices $\bfK_{l}\in \cone_{G_{l}}$, graphs $G_{l}\in \mathcal{G}_{m_{l}}$ and auxiliary variables $z_{l}$ for $1\le l \le L$:
\begin{eqnarray} \label{eq:jointpost}
 \pr\left(\bfK_{1},G_{1},(\bfK_{l},G_{l},z_{l})_{l=2}^{L}\mid \mathcal{D}\right) \propto \pr(\mathcal{D}\mid \bfK_{1},\ldots,\bfK_{L}) \pr(\bfK_{1}\mid G_{1}) \prod_{l=2}^{L} \pr(z_{l},\bfK_{l}\mid G_{l}) \prod_{l=1}^{L} \pi_{m_{l}}(G_{l})
\end{eqnarray}
Here $\pi_{m_{l}}(G_{l})$ are prior probabilities on the set of graphs $\mathcal{G}_{m_{l}}$. The full conditionals of $\bfK_{l}$, $1\le l\le L$ and $z_{l}$, $2\le l\le L$ are $G$-Wishart and Gamma, respectively:
\begin{eqnarray*}
 \pr(\bfK_{l}\mid \mbox{rest}) & = & \Wis_{G_{l}} \left( \frac{mn}{m_{l}}+ \delta_{l},\bfS_{l}+z_{l}D_{l}\right),\\
 \pr(z_{l}\mid \mbox{rest}) &= & \mbox{Gamma} \left( \frac{m_{l}(\delta_{l}-2)}{2}+|\nu\left( G_{l}\right)|, \frac{1}{2}\langle \bfK_{l},\bfD_{l}\rangle\right),
\end{eqnarray*}
\noindent where $\mbox{Gamma}(\alpha,\beta)$ has mean $\alpha/\beta$. We use the approach for updating $\bfK_{l}$ ($1\le l\le L$)  described in \cite{DoLe11}. Their method sequentially perturbs each free element in the Cholesky decomposition of each precision matrix. The constraint (\ref{eq:k11}) is imposed by not updating the free element $(\bfphi_{l})_{11}=\sqrt{(\bfK_{l})_{11}}=1$.\\
\indent The updates of the graphs $G_{l}$ are based on the full joint conditionals of $\bfK_{l}$ and $G_{l}$, $1\le l\le L$:
\begin{eqnarray*}
 \pr (\bfK_{l},G_{l}\mid \mbox{rest}) & \propto & \frac{1}{I_{G_{l}}(\delta_{l},\bfD_{l})} (\mbox{det}\; \bfK_{l})^{\frac{1}{2}\left( \frac{mn}{m_{l}}+\delta_{l}-2\right)}z_{l}^{\frac{m_{l}(\delta_{l}-2)}{2}+|\nu\left( G_{l}\right)|-1}\exp\left\{ -\frac{1}{2}\langle \bfK_{l},\bfS_{l}+z_{l}\bfD_{l}\rangle\right\},
\end{eqnarray*}
\noindent since, once an edge in $G_{l}$ is added or deleted, the corresponding set of free elements of $\bfK_{l}$ together with the remaining bound elements must also be updated.\\
\indent We denote by $\nbd_{m_{l}}^{+}(G_{l})$ the graphs that can be obtained by adding an edge to a graph $G_{l}\in \ground_{m_{l}}$ and by $\nbd_{m_{l}}^{-}(G_{l})$ the graphs that are obtained by deleting an edge from $G_{l}$. We call the one-edge-way set of graphs $\nbd_{m_{l}}(G_{l})=\nbd_{m_{l}}^{+}(G_{l})\cup \nbd_{m_{l}}^{-}(G_{l})$ the neighborhood of $G_{l}$ in $\ground_{m_{l}}$. These neighborhoods connect any two graphs in $\ground_{m_{l}}$ through a sequence of graphs such that two consecutive graphs in this sequence are each others' neighbors. We sample a candidate graph $G^{\prime}_{l}\in \nbd_{m_{l}}(G_{l})$ from the proposal distribution:
\begin{eqnarray} \label{eq:fairproposalcol}
 q\left( G_{l}^{\prime}\mid G_{l},z_{l}\right) = \frac{1}{2}\frac{z_{l}^{|\nu\left( G^{\prime}_{l}\right)|}}{\sum\limits_{G^{\prime\prime}_{l}\in \nbd^{+}_{m_{l}}\left(G_{l}\right)}z_{l}^{|\nu\left( G^{\prime\prime}_{l}\right)|}} \delta_{\left\{ G^{\prime}_{l}\in \nbd^{+}_{m_{l}}\left(G_{l}\right)\right\}} + \frac{1}{2}\frac{z_{l}^{|\nu\left( G^{\prime}_{l}\right)|}}{\sum\limits_{G^{\prime\prime}_{l}\in \nbd^{-}_{m_{l}}\left(G_{l}\right)} z_{l}^{|\nu\left( G^{\prime\prime}_{l}\right)|}} \delta_{\left\{ G^{\prime}_{l}\in \nbd^{-}_{m_{l}}\left(G_{l}\right)\right\}},
\end{eqnarray}
\noindent where $\delta_{A}$ is equal to $1$ if $A$ is true and is $0$ otherwise. The proposal (\ref{eq:fairproposalcol}) gives an equal probability that the candidate graph is obtained by adding or deleting an edge from the current graph $G_{l}$.\\
\indent We assume that the candidate graph $G^{\prime}_{l}$ is obtained by adding an edge $(v_{1},v_{2})$, $v_{1}<v_{2}$, to $G_{l}$. We have $\nu(G^{\prime}_{l})=\nu(G_{l})\cup \{(v_{1},v_{2})\}$, $\bd_{G^{\prime}_{l}}(v_{1}) = \bd_{G_{l}}(v_{1})\cup \{ v_{2}\}$ and $d_{v_{1}}^{G^{\prime}_{l}}=d_{v_{1}}^{G_{^{s}l}}+1$. We define an upper diagonal matrix $\bfphi^{\prime}_{l}$ such that $(\bfphi^{\prime}_{l})_{v_{1}^{\prime},v_{2}^{\prime}}=(\bfphi_{l})_{v_{1}^{\prime},v_{2}^{\prime}}$ for all $(v^{\prime}_{1},v^{\prime}_{2})\in \nu(G_{l})$. The value of $(\bfphi^{\prime}_{l})_{v_{1},v_{2}}$sampled from a $\normal\left((\bfphi_{l})_{v_{1},v_{2}},\sigma_{g}^{2}\right)$ distribution. The bound elements of $\bfphi^{\prime}_{l}$ are determined through completion with respect to $G^{\prime}_{l}$. We form the candidate matrix $\bfK^{\prime}_{l}=(\bfphi^{\prime}_{l})^{T}\bfphi^{\prime}_{l}\in \cone_{G^{\prime}_{l}}$. Since the dimensionality of the parameter space increases by one, we must make use of the reversible jump method of \cite{Gr95}. We accept the update of $(\bfK_{l},G_{l})$ to $(\bfK^{\prime}_{l},G^{\prime}_{l})$ with probability $\min\{R_{g},1\}$, where
\begin{eqnarray*}
 R_{g} = \frac{\pr(\bfK^{\prime}_{l},G^{\prime}_{l}\mid \mbox{rest})}{\pr(\bfK_{l},G_{l}\mid \mbox{rest})}\frac{q\left( G_{l}\mid G^{\prime}_{l},z_{l}\right)}{q\left( G^{\prime}_{l}\mid G_{l},z_{l}\right)} \frac{J\left(\bfK^{\prime}_{l} \rightarrow \bfphi^{\prime}_{l}\right)}{J\left(\bfK_{l} \rightarrow \phi_{l}\right)} \frac{\pi_{m_{l}}(G^{\prime}_{l})}{\pi_{m_{l}}(G_{l})} \frac{J(\bfphi_{l}\rightarrow \bfphi^{\prime}_{l})}{\frac{1}{\sigma_{g}\sqrt{2\pi}}\exp\left( -\frac{\left((\bfphi^{\prime}_{l})_{v_{1},v_{2}}-(\bfphi_{l})_{v_{1},v_{2}}\right)^{2}}{2\sigma_{g}^{2}}\right)}.
\end{eqnarray*}
Since the free elements of $\bfphi^{\prime}_{l}$ are the free elements of $\bfphi_{l}$ and $(\bfphi^{\prime}_{l})_{v_{1},v_{2}}$, the Jacobian of the transformation from $\bfphi_{l}$ to $\bfphi^{\prime}_{l}$ is equal to $1$. Moreover $\mbox{det}\; \bfK^{\prime}_{l}=\mbox{det}\; \bfK_{l}$ and $J\left(\bfK^{\prime}_{l} \rightarrow \bfphi^{\prime}_{l}\right) = (\bfphi_{l})_{v_{1},v_{1}} J\left(\bfK_{l} \rightarrow \bfphi_{l}\right)$.
It follows that
\begin{eqnarray*}
 R_{g} & = & \sigma_{g}\sqrt{2\pi}(\bfphi_{l})_{v_{1},v_{1}}z_{l} \frac{I_{G_{l}}(\delta_{l},\bfD_{l})}{I_{G^{\prime}_{l}}(\delta_{l},\bfD_{l})} \frac{q\left( G_{l}\mid G^{\prime}_{l},z_{l}\right)}{q\left( G^{\prime}_{l}\mid G_{l},z_{l}\right)} \frac{\pi_{m_{l}}(G^{\prime}_{l})}{\pi_{m_{l}}(G_{l})}\times\\ 
  && \times \exp\left\{-\frac{1}{2}\langle \bfK^{\prime}_{l}-\bfK_{l},\bfS_{l}+z_{l}\bfD_{l}\rangle + \frac{\left((\bfphi^{\prime}_{l})_{v_{1},v_{2}}-(\bfphi_{l})_{v_{1},v_{2}}\right)^{2}}{2\sigma_{g}^{2}}\right\}.
\end{eqnarray*}
\noindent Next we assume that  $G^{\prime}_{l}$ is obtained by deleting the edge $(v_{1},v_{2})$ from $G_{l}$. We have $\nu(G^{\prime}_{l})=\nu(G_{l})\setminus \{(v_{1},v_{2})\}$ , $\bd_{G^{\prime}_{l}}(v_{1}) = \bd_{G_{l}}(v_{1})\setminus \{ v_{2}\}$ and $d_{v_{1}}^{G^{\prime}_{l}}=d_{v_{1}}^{G^{\prime}_{l}}-1$. We define an upper diagonal matrix $\bfphi^{\prime}_{l}$ such that $(\bfphi^{\prime}_{l})_{v_{1}^{\prime},v_{2}^{\prime}}=(\bfphi_{l})_{v_{1}^{\prime},v_{2}^{\prime}}$ for all $(v^{\prime}_{1},v^{\prime}_{2})\in \nu(G^{\prime}_{l})$. The bound elements of $\bfphi^{\prime}_{l}$ are obtained by completion with respect to $G^{\prime}_{l}$. The candidate precision matrix is $\bfK^{\prime}_{l}=(\bfphi^{\prime}_{l})^{T}\bfphi^{\prime}_{l}\in \cone_{G^{\prime}_{l}}$. Since the dimensionality of the parameter space decreases by $1$, the acceptance probability of the update of $(\bfK_{l},G_{l})$ to $(\bfK^{\prime}_{l},G^{\prime}_{l})$ is $\min\{R_{g}^{\prime},1\}$, where
\begin{eqnarray*}
 R^{\prime}_{g}  & = & \left(\sigma_{g}\sqrt{2\pi}(\bfphi_{l})_{v_{1},v_{1}}z_{l}\right)^{-1} \frac{I_{G_{l}}(\delta_{l},\bfD_{l})}{I_{G^{\prime}_{l}}(\delta_{l},\bfD_{l})} \frac{q\left( G_{l}\mid G^{\prime}_{l},z_{l}\right)}{q\left( G^{\prime}_{l}\mid G_{l},z_{l}\right)}\frac{\pi_{m_{l}}(G^{\prime}_{l})}{\pi_{m_{l}}(G_{l})}\times\\
 \\ && \times \exp\left\{-\frac{1}{2}\langle \bfK^{\prime}_{l}-\bfK_{l},\bfS_{l}+z_{l}\bfD_{l}\rangle - \frac{\left((\bfphi^{\prime}_{l})_{v_{1},v_{2}}-(\bfphi_{l})_{v_{1},v_{2}}\right)^{2}}{2\sigma_{g}^{2}}\right\}.
\end{eqnarray*}

\section{Multi-way GGMs with Separable Mean Parameters} \label{sec:sepmeans}

So far we have discussed multi-way GGMs associated with array normal distributions with a $m_{1}\times\ldots\times m_{L}$ array mean parameter $\bfM$ assumed to be zero. In some practical applications this assumption is too restrictive and $\bfM$ needs to be explicitly accounted for. The observed samples $\mathcal{D}=\{\bfx^{(1)},\ldots,\bfx^{(n)}\}$ grouped as an $m_{1}\times\ldots\times m_{L}\times n$ array are modeled as
\begin{eqnarray}\label{eq:sepmeans}
 \mathcal{D} = \bfM\circ \mathbf{1}_{n} + \bfX,\quad \bfX \sim \anormal_{L+1}(\mathbf{0};\{m_{1},\bfK_{1}\},\ldots,\{m_{L},\bfK_{L}\},\{ n,\bfK_{L+1}\}), \quad \bfK_{L+1}= \bfI_{n}.
\end{eqnarray}
If the sample size $n$ is small or if the observed samples are not independent and their dependence structure is represented by removing the constraint $\bfK_{L+1}=\bfI_{n}$, estimating $m=\prod_{l=1}^{L}m_{l}$ mean parameters is unrealistic. The matrix-variate normal models of \cite{allen-tibshirani-2010} have separable mean parameters defined by row and column means, while \cite{allen-2011} extends separable means for general array data. The $L$-dimensional mean array $\bfM$ is written as a sum of distinct mean arrays associated with each dimension:
$$
 \bfM = \sum_{l=1}^{L} \bfM_{l}, \mbox{ where } \bfM_{l} = \mathbf{1}_{m_{1}}\circ \ldots \circ \mathbf{1}_{m_{l-1}}\circ \bfmu_{l}\circ \mathbf{1}_{m_{l+1}}\circ \ldots \circ \mathbf{1}_{m_{L}}, 1\le l\le L.
$$
Here $\bfmu_{l}\in \reals^{m_{l}}$ represents the mean vector associated with dimension $l$ of $\bfX$. This particular structure of the mean array $\bfM$ implies the following marginal distribution for each element of the array of random effects $\bfX$:
$$
 X_{i_{1}\ldots i_{L}i_{L+1}} \sim \normal\left( \sum_{l=1}^{L}(\bfmu_{l})_{i_{l}}, \prod_{l=1}^{L} (\bfK_{l}^{-1})_{i_{l}i_{l}}\right).
$$
Thus $\bfmu_{l}$ can be interpreted as fixed effects associated with dimension $l$. The dependency between two different elements of $\bfX$ is represented by their covariance
$$
 \Cov(X_{i_{1}\ldots i_{L}i_{L+1}},X_{i^{\prime}_{1}\ldots i^{\prime}_{L}i^{\prime}_{L+1}}) =  \prod_{l=1}^{L} (\bfK_{l}^{-1})_{i_{l}i^{\prime}_{l}}.
$$
We remark that the individual mean arrays $\bfM_{l}$ are not identified, but their sum $\bfM$ is identified.\\
\indent Bayesian estimation of $\bfM$ proceeds by specifying independent priors $\bfmu_{l}\sim \normal_{m_{l}}(\bfmu^{0}_{l},\bfOmega_{l}^{-1})$. To simplify the notations we take $n=1$, hence the arrays in equation (\ref{eq:sepmeans}) have only $L$ dimensions. We develop a Gibbs sampler in which each vector $\bfmu_{l}$ is updated as follows. Denote $m_{-l} = \prod_{l^{\prime}\ne l}m_{l^{\prime}}$ and consider the $l$-mode matricizations $\mathcal{D}_{(l)}$, $\bfM_{(l)}$ and $\bfX_{(l)}$ of the arrays $\mathcal{D}$, $\bfM$ and $\bfX$. From equation (\ref{eq:sepmeans}) it follows that the $m_{l}\times m_{-l}$ random matrix
$$
 \widetilde{\bfX}_{(l)} = \bfX_{(l)}-\sum_{l^{\prime}\ne l} (\bfM_{l^{\prime}})_{(l)}
$$
follows a matrix-variate normal distribution with mean $(\bfM_{l})_{(l)} = \mu_{l}1^{T}_{m_{-l}}$, row precision matrix $\bfK_{l}$ and column precision matrix $\bfK_{-l} = \bfK_{L}\otimes\ldots\otimes \bfK_{l+1}\otimes \bfK_{l-1} \otimes\ldots\otimes \bfK_{1}$. It follows that $\bfmu_{l}$ is updated by direct sampling from the multivariate normal $\normal_{m_{l}} \left( \mathbf{m}_{\mu_{l}}, K_{\mu_{l}}^{-1}\right)$ where
\begin{eqnarray*}
 \bfK_{\mu_{l}} = \left( \mathbf{1}_{m_{-l}}^{T} \bfK_{-l} \mathbf{1}_{m_{-l}}\right) \bfK_{l} + m_{-l}\bfOmega_{l},\quad
 \mathbf{m}_{\mu_{l}} = \bfK_{\mu_{l}}^{-1}\left[ \bfK_{l} \widetilde{\bfX}_{(l)} \bfK_{-l}\mathbf{1}_{m_{-l}} + m_{-l}\bfOmega_{l}\bfmu^{0}_{l} \right].
\end{eqnarray*}

\section{An Application to Spatiotemporal Cancer Mortality Surveillance} \label{sec:poissonmodel}

In this section we construct a spatial hierarchical model for spatiotemporal cancer mortality surveillance based on the multi-way GGMs just developed.  A relevant dataset could comprise counts $y_{i,j,t}$ for the number of deaths from cancer $i$ in area $j$ on year $t$, and can be seen as a three-dimensional array of size $m_{C} \times m_{S} \times m_{T}$. Our proposed model accounts for temporal and spatial dependence in mortality counts, as well as dependence across cancer types:
\begin{align*}
y_{i_{C},i_{S},i_{T}} \mid \theta_{i_{C},i_{S},i_{T}} \sim \Poi\left( \exp \left\{ \mu_{i_{C}} + \log( h_{i_{S},i_{T}})  + \theta_{i_{C},i_{S},i_{T}}  \right\} \right).
\end{align*}
Here $i_{C}=1,\ldots,m_{C}, \; i_{S}=1,\ldots,m_{S}, \; i_{T}=1,\ldots,m_{T}$, $h_{i_{S},i_{T}}$ denotes the population in area $i_{S}$ during year $i_{T}$, $\mu_{i_{C}}$ is the mean number of deaths due to cancer $i_{C}$ over the whole period and all locations, and $\theta_{i_{C},i_{S},i_{T}}$ is a zero-mean random effect, which is assigned the prior:
\begin{align*}
\bfTheta = (\theta_{i_{C},i_{S},i_{T}})& \sim \anormal_{3}( \mathbf{0};\{ m_{C}, \bfK_{C} \}, \{ m_{S}, \bfK_{S} \}, \{ m_{T}, \bfK_{T} \} ).
\end{align*}
The matrix $\bfK_{C}$ models dependence across cancer types, $\bfK_{S}$ accounts for spatial dependence across neighboring areas, and $\bfK_{T}$ accounts for temporal dependence.  Since we do not have prior information about dependence across cancer types, the prior for $\bfK_{C}$ is specified hierarchically by setting
\begin{align*}
\bfK_{C} \mid G_{C} &\sim \Wis_{G_{C}} (\delta_{C}, \mathbf{I}_{m_{C}}), & \pr(G_{C}) & \propto 1. 
\end{align*}
Thus $G_{C}\in \mathcal{G}_{m_{C}}$ defines the unknown graphical model of cancer types. For the spatial component, we follow the approach of \cite{DoLeRo11} and use a GGM to specify a conditionally autoregressive prior
\begin{align*}
(z_{S}\bfK_{S}) \mid G_{S} &\sim \Wis_{G_{S}} (\delta_{S}, (\delta_{S}-2)\bfD_{S}), 
\end{align*}
where $\bfD_{S} = (\bfE_{\bfW} - \rho \bfW)^{-1}$ and $W$ is the adjacency matrix for the $m_{S}$ areas, so that $W_{i^{1}_{S},i^{2}_{S}} = 1$ if areas $i^{1}_{S}$ and $i^{2}_{S}$ share a common border, and $W_{i^{1}_{S},i^{2}_{S}}=0$ otherwise, and $\bfE_{\bfW} =\diag \{ \mathbf{1}^{T}_{m_{S}}\bfW \}$.  The graph $G_{S}$ is fixed and given by the adjacency matrix $\bfW$. Furthermore, we assume that, {\it a priori}, there is a strong degree of positive spatial association, and choose a prior for spatial autocorrelation parameter  $\rho$ that gives higher probabilities to values close to $1$ (see \cite{GeVo03}):
\begin{eqnarray*}
 \rho & \sim & \Uni(\{ 0,0.05,0.1,\ldots,0.8,0.82,\ldots,0.90,0.91,\ldots,0.99\}).
\end{eqnarray*}
For the temporal component, the prior for $\bfK_{T}$ is set to
\begin{align*}
(z_{T}\bfK_{T}) \mid G_{T} &\sim \Wis_{G_{T}} (\delta_{T}, \mathbf{I}_{m_{T}}).
\end{align*}
The graph $G_{T}$ gives the temporal pattern of dependence and could be modeled in a manner similar to the graph $G_C$ for cancer types. Instead of allowing $G_{T}$ to be any graph with $m_T$ vertices, we can constrain it to belong to a restricted set of graphs, for example, the graphs $G_{T}^{(1)}$, $G_{T}^{(2)}$, $G_{T}^{(3)}$ and $G_{T}^{(4)}$ with vertices $\{1,2,\ldots,m_{T}\}$ and edges $E_{T}^{(1)}$, $E_{T}^{(2)}$, $E_{T}^{(3)}$ and $E_{T}^{(4)}$, where
\begin{eqnarray*}
 E_{T}^{(1)} & = & \{ (i_{T}-1,i_{T}) : 2\le i_{T}\le m_{T}\},\\
 E_{T}^{(2)} & = & E_{T}^{(1)}  \cup\{ (i_{T}-2,i_{T}) : 3\le i_{T}\le m_{T}\},\\
 E_{T}^{(3)} & = & E_{T}^{(2)}  \cup\{ (i_{T}-3,i_{T}) : 4\le i_{T}\le m_{T}\},\\
 E_{T}^{(4)} & = & E_{T}^{(3)}  \cup\{ (i_{T}-4,i_{T}) : 5\le i_{T}\le m_{T}\}.
\end{eqnarray*}
These four graphs define AR(1), AR(2), AR(3) and AR(4) models. We set $\delta_{C}=\delta_{S}=\delta_{T}=3$. We use a multivariate normal prior for the mean rates vector $\bfmu=(\mu_{1},\ldots,\mu_{m_{C}})^{T} \sim \normal_{m_{C}}(\bfmu^{0},\bfOmega^{-1})$ where $\bfmu^{0}= \mu_{0} \mathbf{1}_{m_{C}}$ and $\bfOmega = \omega^{-2}\mathbf{I}_{m_{C}}$. We set $\mu_{0}$ to be the median log incidence rate across all cancers, all areas and all time points, and $\omega$ to be twice the interquartile range of the raw log incidence rates.\\
\indent The MCMC algorithm for this sparse multivariate spatiotemporal model involves iterative updates of the precision matrices $\bfK_{C}$, $\bfK_{S}$ and $\bfK_{T}$ as well as of the graph $G_{C}$ as described in Section \ref{sec:mcmc}. The mean rates $\bfmu$ are sampled as described in Section \ref{sec:sepmeans}. Here the three-dimensional mean parameter array $\bfM$ is equal with the mean array associated with the first dimension (cancers), while the mean arrays associated with the other two dimensions (space and time) are set to zero:
$$
 \bfM = \bfmu \circ \mathbf{1}_{m_{S}} \circ \mathbf{1}_{m_{T}}.
$$
We consider the centered random effects $\widetilde{\bfTheta} = \bfM + \bfTheta$  which follows an array normal distribution with mean $\bfM$ and precision matrices $\bfK_{C}$, $\bfK_{S}$, $\bfK_{T}$. We form $\widetilde{\bfTheta}_{(1)}$ \---- the 1-mode matricization of $\widetilde{\bfTheta}$. It follows that $\bar{\bfTheta} = \widetilde{\bfTheta}_{(1)}^{T}$ is a $(m_{S}m_{T})\times m_{C}$ matrix that follows a matrix-variate normal distribution with mean $\mathbf{1}_{m_{S}m_{T}}\mu^{T}$, row precision matrix $\bar{\bfK}_{R} = \bfK_{T}\otimes \bfK_{S}$ and column precision matrix $\bfK_{C}$. We resample $\widetilde{\bfTheta}$ by sequentially updating each row vector $\bar{\bfTheta}_{i\textasteriskcentered}$, $i=1,\ldots,m_{S}m_{T}$. Conditional on the other rows of $\bar{\bfTheta}$, the distribution of $\left(\bar{\bfTheta}_{i\textasteriskcentered}\right)^{T}$ with $i=i_{S}i_{T}$ ($1\le i_{S}\le m_{S}$, $1\le i_{T}\le m_{T}$) is multivariate normal with mean $\bfM_{i}$ and precision matrix $\bfV_{i}$, where
$$
 \bfM_{i} = \bfmu - \sum_{i^{\prime}=1}^{m_{S}m_{T}} \frac{(\bar{\bfK}_R)_{ii^{\prime}}}{(\bar{\bfK}_R)_{ii}} \left[\left(\bar{\bfTheta}_{i^{\prime}\textasteriskcentered}\right)^{T} -\mu \right],\quad \bfV_{i} = \left( \bar{\bfK}_{R}\right)_{ii}\bfK_{C}.
$$
Thus the full conditional distribution of $\bar{\bfTheta}_{i\textasteriskcentered}$ is proportional with
\begin{eqnarray}\label{eq:fullcond_step1}
 & \prod\limits_{i_{C}=1}^{m_{C}}  \exp\left\{  y_{i_{C},i_{S},i_{T}}\left( \mu_{i_{C}} + \log( h_{i_{S},i_{T}})  + \theta_{i_{C},i_{S},i_{T}} \right) - h_{i_{S},i_{T}} \exp\left( \mu_{i_{C}} + \theta_{i_{C},i_{S},i_{T}} \right) \right\} \times & \nonumber \\
  &\times \exp\left\{  -\frac{1}{2} \left[ \bar{\bfTheta}_{i\textasteriskcentered} - \left( \bfM_{i}\right)^{T}\right] \bfV_{i}  \left[ \left(\bar{\bfTheta}_{i\textasteriskcentered}\right)^{T} - \bfM_{i}\right] \right\}. &
\end{eqnarray}
We make use of a Metropolis-Hastings step to sample from to sample from (\ref{eq:fullcond_step1}).  We consider a strictly positive precision parameter $\widetilde{\sigma}$. For each $i_{C}=1,\ldots,m_{C}$, we update the $i_{C}$-th element of $\bar{\bfTheta}_{i\textasteriskcentered}$ by sampling $\gamma\sim \normal\left( \bar{\Theta}_{i,i_{C}},\widetilde{\sigma}^{2}\right)$. We define a candidate row vector $\bar{\bfTheta}^{new}_{i\textasteriskcentered}$ by replacing $\bar{\Theta}_{i,i_{C}}$ with $\gamma$ in $\bar{\bfTheta}_{i\textasteriskcentered}$. We update the current $i$-th row of $\bar{\bfTheta}$ with $\bar{\bfTheta}^{new}_{i\textasteriskcentered}$ with the Metropolis-Hastings acceptance probability corresponding with (\ref{eq:fullcond_step1}). Otherwise the $i$-th row of $\bar{\bfTheta}$ remains unchanged.

\section{Dynamic Multi-way GGMs for Array-variate Time Series} \label{sec:dynamic}
 
The cancer mortality surveillance application from Section \ref{sec:poissonmodel} represented the time component as one of the dimensions of the three-dimensional array of observed counts. We give an extension of multi-way GGMs to array-variate time series $\bfY_{t}$, $t=1,2,\ldots,T$, where $\bfY_{t}\in \reals^{m_{1}\times\ldots\times m_{L}}$. Our framework generalizes the results from \cite{carvalho-west-2007a,carvalho-west-2007b} and \cite{wang_west_2009} which assume vector ($L=1$) or matrix-variate ($L=2$) time series. We build on the standard specification of Bayesian dynamic linear models \cite{WeHa97}, and assume that $\bfY_{t}$ is modeled over time by
\begin{eqnarray} 
 \bfY_{t} & = & \bfTheta_{t} \times_{L+1} \mathbf{F}_{t}^{T} + \mathbf{\Psi}_{t}, \quad \mathbf{\Psi}_{t}\sim\anormal_{L}(\mathbf{0};\{m_{1},v_{t}^{-1}\bfK_{1}\},\{m_{2},\bfK_{2}\},\ldots,\{m_{L},\bfK_{L}\}),\label{eq:dlmobseq}\\
 \bfTheta_{t} & = & \bfTheta_{t-1}\times_{L+1} \mathbf{H}_{t} + \mathbf{\Gamma}_{t}, \quad \mathbf{\Gamma}_{t}\sim \anormal_{L+1}(\mathbf{0};\{m_{1},\bfK_{1}\},\ldots,\{m_{L},\bfK_{L}\},\{s,\bfW_{t}^{-1}\}),\label{eq:dlmeveq}
\end{eqnarray}
\noindent where (a) $\bfTheta_{t}\in  \reals^{m_{1}\times\ldots\times m_{L}\times s}$ is the state array at time $t$; (b) $\mathbf{F}_{t}\in \reals^{s}$ is a vector of known regressors at time $t$; (c) $\mathbf{\Psi}_{t}\in \reals^{m_{1}\times\ldots\times m_{L}}$ is the array of observational errors at time $t$; (d) $\mathbf{H}_{t}$ is a known $s\times s$ state evolution matrix at time $t$; (e) $\mathbf{\Gamma}_{t}\in  \reals^{m_{1}\times\ldots\times m_{L}\times s}$ is the array of state evolution innovations at time $t$; (f) $\bfW_{t}$ is the $s\times s$ innovation covariance matrix at time $t$; (g) $v_{t}>0$ is a known scale factor at time $t$. Furthermore, the observational errors $\mathbf{\Psi}_{t}$ and the state evolution errors $\mathbf{\Gamma}_{t}$ follow zero-mean array normal distributions defined by $\bfK_{1},\ldots,\bfK_{L}$ and $\bfW_{t}$, and are assumed to be both independent over time element-wise and mutually independent as sequences of arrays.\\
\indent The observation equation (\ref{eq:dlmobseq}) and the evolution equation (\ref{eq:dlmeveq}) translate into the following dynamic linear model for the univariate time series $(\bfY_{t})_{i_{1}\ldots i_{L}}$, $t=1,2,\ldots,T$:
\begin{eqnarray*}
 (\bfY_{t})_{i_{1}\ldots i_{L}} & = & \mathbf{F}_{t}^{T} (\bfTheta_{t})_{i_{1}\ldots i_{L}\star} + (\mathbf{\Psi}_{t})_{i_{1}\ldots i_{L}}, \quad  (\mathbf{\Psi}_{t})_{i_{1}\ldots i_{L}}\sim \normal\left(0,v_{t}\prod_{l=1}^{L}(\bfK^{-1}_{l})_{i_{l}i_{l}}\right),\\
 (\bfTheta_{t})_{i_{1}\ldots i_{L},\star} & = & \mathbf{H}_{t} (\bfTheta_{t-1})_{i_{1}\ldots i_{L},\star} + (\mathbf{\Gamma}_{t})_{i_{1}\ldots i_{L},\star},\quad (\mathbf{\Gamma}_{t})_{i_{1}\ldots i_{L},\star}\sim \normal_{s}\left( \mathbf{0},\prod_{l=1}^{L}(\bfK^{-1}_{l})_{i_{l}i_{l}} \bfW_{t}\right),
\end{eqnarray*}
\noindent where $(\bfTheta_{t})_{i_{1}\ldots i_{L},\star}= ((\bfTheta_{t})_{i_{1}\ldots i_{L}1},\ldots,(\bfTheta_{t})_{i_{1}\ldots i_{L}s})^{T}$, while $(\bfTheta_{t-1})_{i_{1}\ldots i_{L},\star}$ and $(\mathbf{\Gamma}_{t})_{i_{1}\ldots i_{L},\star}$ are defined in a similar manner. The components $\mathbf{F}_{t}$, $\mathbf{H}_{t}$ and $\bfW_{t}$ are the same for all univariate time series, but the state parameters $(\bfTheta_{t})_{i_{1} \ldots,i_{L} \star}$ as well as their scales of measurement defined by $\prod_{l=1}^{L}(\bfK^{-1}_{l})_{i_{l}i_{l}}$ could be different across series. The cross-sectional dependence structure across individual time series at time $t$ is induced by $\bfK_{1},\ldots,\bfK_{L}$ and $\bfW_{t}$:
\begin{eqnarray*}
 \Cov\left( (\mathbf{\nu}_{t})_{i_{1}\ldots i_{L}}, (\mathbf{\nu}_{t})_{i^{\prime}_{1}\ldots i^{\prime}_{L}}\right) =  v_{t}\prod_{l=1}^{L} (\bfK_{l}^{-1})_{i_{l}i^{\prime}_{l}}, \quad \Cov\left( (\mathbf{\Gamma}_{t})_{i_{1}\ldots i_{L},\star}, (\mathbf{\Gamma}_{t})_{i^{\prime}_{1}\ldots i^{\prime}_{L},\star}\right) =  \prod_{l=1}^{L} (\bfK_{l}^{-1})_{i_{l}i^{\prime}_{l}} \bfW_{t}.
\end{eqnarray*}
For example, if $\prod_{l=1}^{L} (\bfK_{l}^{-1})_{i_{l}i^{\prime}_{l}}$ is large in absolute value, the univariate time series $(\bfY_{t})_{i_{1}\ldots i_{L}}$ and $(\bfY_{t})_{i^{\prime}_{1}\ldots i^{\prime}_{L}}$ exhibit significant dependence in the variation of their observational errors and state vectors. Appropriate choices for the matrix sequence $\bfW_{t}$, $t=1,2,\ldots,T$, arise from the discount factors discussed in \cite{WeHa97} as exemplified, among others, in \cite{wang_west_2009}. The scale factors $v_{t}$ can be set to $1$, but other suitable values can be employed as needed.\\
\indent The following result extends Theorem 1 of \cite{wang_west_2009} to array-variate time series.
\begin{theorem} \label{th:ggmdlms}
 Let $\mathcal{D}_{0}$ be the prior information and denote by $\mathcal{D}_{t} = \{\bfY_{t},\mathcal{D}_{t-1}\}$ the information available at time $t=1,2,\ldots,T$. We assume to have specified precision matrices $\bfK_{1},\ldots,\bfK_{L}$, the matrix sequence $\bfW_{t}$, $t=1,2,\ldots,T$, as well as an initial prior for the state array at time $0$,
$$
 (\bfTheta_{0}\mid \mathcal{D}_{0}) \sim \anormal_{L+1}(\mathbf{M}_{0};\{m_{1},\bfK_{1}\},\ldots,\{m_{L},\bfK_{L}\},\{ s,\mathbf{C}^{-1}_{0}\}),
$$
\noindent where $\mathbf{M}_{0} \in \reals^{m_{1}\times\ldots\times m_{L}\times s}$ and $\mathbf{C}_{0}$ is an $s\times s$ covariance matrix. For every $t=1,2,\ldots,T$, the following distributional results hold:\\
(i) posterior at $t-1$:
$$
 (\bfTheta_{t-1}\mid \mathcal{D}_{t-1}) \sim \anormal_{L+1}(\mathbf{M}_{t-1};\{m_{1},\bfK_{1}\},\ldots,\{m_{L},\bfK_{L}\},\{ s,\mathbf{C}^{-1}_{t-1}\}),
$$
(ii) prior at $t$:
$$
 (\bfTheta_{t}\mid \mathcal{D}_{t-1}) \sim \anormal_{L+1}(\mathbf{a}_{t};\{m_{1},\bfK_{1}\},\ldots,\{m_{L},\bfK_{L}\},\{ s,\mathbf{R}^{-1}_{t}\}),
$$
\noindent where $\mathbf{a}_{t} = \mathbf{M}_{t-1} \times_{L+1} \mathbf{H}_{t}$ and $\mathbf{R}_{t} = \mathbf{H}_{t}\mathbf{C}_{t-1}\mathbf{H}_{t}^{T}+\bfW_{t}$.\\
(iii) one-step forecast at $t-1$:
$$
 (\bfY_{t}\mid \mathcal{D}_{t-1}) \sim \anormal_{L}(\mathbf{f}_{t};\{m_{1},q_{t}^{-1}\bfK_{1}\},\{m_{2},\bfK_{2}\},\ldots,\{m_{L},\bfK_{L}\}),
$$
\noindent where $\mathbf{f}_{t} = \mathbf{M}_{t-1}\times_{L+1} (\mathbf{F}_{t}^{T}\mathbf{H}_{t}) = \mathbf{a}_{t}\times_{L+1} \mathbf{F}_{t}^{T}$ and $q_{t} = \mathbf{F}_{t}^{T}\mathbf{R}_{t} \mathbf{F}_{t}+v_{t}$.\\
(iv) posterior at $t$:
$$
 (\bfTheta_{t}\mid \mathcal{D}_{t}) \sim \anormal_{L+1}(\mathbf{M}_{t};\{m_{1},\bfK_{1}\},\ldots,\{m_{L},\bfK_{L}\},\{ s,\mathbf{C}^{-1}_{t}\}),
$$
\noindent where $\mathbf{M}_{t} = \mathbf{a}_{t}+\mathbf{e}_{t} \times_{L+1} \mathbf{A}_{t}$, $\mathbf{C}_{t} = \mathbf{R}_{t}-\mathbf{A}_{t}\mathbf{A}_{t}^{T}q_{t}$. Here $\mathbf{A}_{t} = q_{t}^{-1}\mathbf{R}_{t}\mathbf{F}_{t}$ and $\mathbf{e}_{t} = \bfY_{t}-\mathbf{f}_{t}$.
\end{theorem}

The proof of Theorem \ref{th:ggmdlms} is straightforward. We write equations (\ref{eq:dlmobseq}) and (\ref{eq:dlmeveq}) in matrix form:
\begin{eqnarray}
 \left( \mathbf{Y}_{t}\right)_{(L+1)} & = & \mathbf{F}^{T}_{t}\left( \mathbf{\Theta}_{t}\right)_{(L+1)} + \left( \mathbf{\Psi}_{t}\right)_{(L+1)}, \quad \left( \mathbf{\Psi}_{t}\right)_{(L+1)}\sim \normal_{m} \left( \mathbf{0}, v_{t}\mathbf{K}^{-1}\right), \label{eq:obsmat}\\
 \left( \mathbf{\Theta}_{t}\right)_{(L+1)}  & = & \mathbf{H}_{t}\left( \mathbf{\Theta}_{t-1}\right)_{(L+1)} + \left( \mathbf{\Psi}_{t}\right)_{(L+1)}, \quad \left( \mathbf{\Psi}_{t}\right)_{(L+1)}\sim \anormal_{2}\left( \mathbf{0}; \{ s,\mathbf{W}_{t}^{-1}\}, \{ m,\mathbf{K}\}\right), \label{eq:evmat}
\end{eqnarray}
where $m=\prod_{l=1}^{L}m_{l}$ and $\mathbf{K}$ is given in equation (\ref{eq:sepnorm}). The normal theory results laid out in \cite{WeHa97} apply directly to the dynamic linear model specified by equations (\ref{eq:obsmat}) and (\ref{eq:evmat}). The predictive distributions relevant for forecasting and retrospective sampling for array-variate time series can be derived from the corresponding predictive distributions for vector data.\\
\indent We complete the definition and prior specification for the dynamic multi-way GGMs with independent G-Wishart priors from equation (\ref{eq:sepnormprior}) for the precision matrices $\mathbf{K}_{1},\ldots,\mathbf{K}_{L}$ and their corresponding auxiliary variables $z_{2},\ldots,z_{L}$. The graphs $G_{1},\ldots,G_{L}$ associated with the G-Wishart priors receive independent priors $\pi_{m_{l}}(G_{l})$ on $\mathcal{G}_{m_{l}}$, $l=1,\ldots,L$. Posterior inference in this framework can be achieved with the following MCMC algorithm that sequentially performs the following steps:

\noindent (A) {\it Resampling the precision matrices, graphs and auxiliary variables}. By marginalizing over the state arrays $\mathbf{\Theta}_{1},\ldots,\mathbf{\Theta}_{L}$, we obtain the marginal likelihood \cite{carvalho-west-2007a,carvalho-west-2007b}:
\begin{eqnarray*}
 \pr\left( \mathbf{Y}_{1},\ldots,\mathbf{Y}_{T}\mid \bfK_{1},G_{1},(\bfK_{l},G_{l},z_{l})_{l=2}^{L}\right) = \prod_{t=1}^{T} \pr\left( \mathbf{Y}_{t} \mid \mathcal{D}_{t-1}, \bfK_{1},G_{1},(\bfK_{l},G_{l},z_{l})_{l=2}^{L}\right).
\end{eqnarray*}
The one-step forecast distribution (iii) from Theorem \ref{th:ggmdlms} implies that 
$$
 \left(q_{t}^{-1/2}(\bfY_{t}-\mathbf{f}_{t})\mid \mathcal{D}_{t-1}\right) \sim \anormal_{L}(\mathbf{0};\{m_{1},\bfK_{1}\},\{m_{2},\bfK_{2}\},\ldots,\{m_{L},\bfK_{L}\}).
$$
We use the filtering equations from Theorem \ref{th:ggmdlms} to produce the centered and scaled array data $\bar{\mathcal{D}}=\{ q_{t}^{-1/2}(\bfY_{t}-\mathbf{f}_{t}):t=1,\ldots,T\}$. Since the elements of $\bar{\mathcal{D}}$ are independent and identically distributed, we update each precision matrix $\bfK_{l}$, graph $G_{l}$ and auxiliary variable $z_{l}$ as described in Section \ref{sec:mcmc} based on $\bar{\mathcal{D}}$.
 
\noindent (B) {\it Resampling the state arrays}. We employ the forward filtering backward algorithm (FFBS) proposed by \cite{carter-kohn-1994,fruhwirth-1994}. Given the current sampled precision matrices, we start by sampling $\mathbf{\Theta}_{T}$ given $\mathcal{D}_{T}$ from the posterior distribution given in (iv) of Theorem \ref{th:ggmdlms}. Then, for $t=T-1,T-2,\ldots,0$, we sample $\mathbf{\Theta}_{t}$ given $\mathcal{D}_{T}$ and $\mathbf{\Theta}_{t+1}$ from the array normal distribution
$$\anormal_{L+1}\left( \mathbf{M}_{t}^{*}; \{ m_{1},\bfK_{1}\},\ldots,\{ m_{L},\bfK_{L}\}, \{ s,(\mathbf{C}^{*}_{t})^{-1}\}\right),$$
where
$$
 \mathbf{M}_{t}^{*} = \mathbf{M}_{t} + \left( \mathbf{\Theta}_{t+1}-\mathbf{a}_{t+1}\right) \times_{L+1} \left( \mathbf{C}_{t}\mathbf{G}^{T}_{t+1}\mathbf{R}^{-1}_{t+1}\right).
$$

\section{Discussion} \label{sec:discussion}

Recent advances in data collection techniques have allowed the creation of high-dimensional public health datasets that monitor the incidence of many diseases across several areas, time points and additional ecological sociodemographic groupings \cite{elliott-et-2001}. Jointly modeling the disease risk associated with each resulting cell count (i.e., a particular disease at a particular time point in a particular region given a particular combination of risk factors) is desirable since it takes into consideration interaction patterns that arise within each dimension or across dimensions. By aggregating data across time, key epidemiological issues related to the evolution of the risk patterns across time might not be given an appropriate answer \cite{abellan-et-2008}. The spatial structure of geographical regions must also be properly accounted for \cite{Be74,besag-york-mollie-1991}. Furthermore, since diseases are potentially related and share risk factors, it is critical that individual models should not be developed for each disease \cite{GeVo03,wang-wall-2003}. Rich, flexible classes of models that capture the joint variation of disease risk in the actual observed data without requiring the aggregation across one or more dimensions will be the fundamental aim of our proposed work related to disease mapping. Multi-way GGMs can be used in Bayesian hierarchical models that produce estimates of disease risk by borrowing strength across time, areas and the other dimensions. Due to the likely presence of small counts in many cells, the degree of smoothing will be controlled through a wide range of parameters that could be constrained to zero according to pre-defined interaction structures (e.g., the neighborhood structure of the areas) or by graphs that received the most support given the data.\\
\indent We generalize the models from Section \ref{sec:poissonmodel}, and let $Y$ be the $L$-dimensional array of observed disease counts indexed by cells $\{ (i_{1},\ldots,i_{L}):1\le i_{l}\le m_{l}\}$. We assume that the count random variable $Y_{i_{1}\ldots i_{L}}$ associated with cell $(i_{1},\ldots,i_{L})$ follows a distribution from an exponential family (e.g., Poisson or binomial) with mean parameter $\theta_{i_{1}\ldots i_{L}}$, i.e.
\begin{eqnarray}\label{eq:expfam}
 Y_{i_{1}\ldots i_{L}}\mid \theta_{i_{1}\ldots i_{L}} & \overset{iid}{\sim} & H(\theta_{i_{1}\ldots i_{L}}),\mbox{ for } 1\le i_{l}\le m_{l}, 1\le l\le L.
\end{eqnarray}
We assume that the cell counts $Y$ are conditionally independent given the $L$-dimensional array of parameters $\theta = \{ \theta_{i_{1}\ldots i_{L}}: 1\le i_{l}\le m_{l}\}$. Furthermore, given a certain link function $g(\cdot)$ (e.g., $\log(\cdot)$), the parameters $\theta$ follow a joint model
\begin{eqnarray} \label{eq:glm}
 g(\theta_{i_{1}\ldots i_{L}}) = \nu_{i_{1}\ldots i_{L}} + X_{i_{1}\ldots i_{L}},
\end{eqnarray}
\noindent where $\nu_{i_{1}\ldots i_{L}}$ is a known offset, while $X=\{X_{i_{1}\ldots i_{L}}:1\le i_{l}\le m_{l}\}$ is an array of zero-centered random effects. Equation (\ref{eq:glm}) can subsequently include explanatory ecological covariates as needed. The multi-way GGMs are employed in the context of non-Gaussian data as joint distribution for the array of random effects $X$. Thus $X$ is assumed to follow the flexible joint distributions, and Each dimension of the data is represented as a GGM in a particular dimension of the random effects $X$.\\
\indent This framework accommodates many types of interactions by restricting the set of graphs that are allowed to represent the dependency patterns of the corresponding dimensions. For example, if dimension $l^{\prime}$ of $Z$ represents time, then the graphs associated with this dimension could be constrained to represent an autoregressive model AR($q$), where $q=1,2,3,\ldots$ \--- see Section \ref{sec:poissonmodel}. Temporal dependence can also be modeled with the dynamic multi-way GGMs from Section \ref{sec:dynamic}. If dimension $l^{\prime\prime}$ represents spatial dependence, one could constrain the space of graphs for dimension $l^{\prime\prime}$ to consist of only one graph with edges defined by areas that are neighbors of each other in the spirit of \cite{Be74,clayton-kaldor-1987}. As opposed to a modeling framework based on GMRFs, we can allow uncertainty around this neighborhood graph in which case we let the space of graphs for dimension $l^{\prime\prime}$ to include graphs that are obtained by adding or deleting one, two or more edges from the neighborhood graph. This expansion of the set of spatial graphs is consistent with the hypothesis that interaction occurs not only between areas that are close to each other or share a border, but also between more distant areas. We can also allow all possible graphs to be associated with dimension $l^{\prime\prime}$ and examine the graphs that receive the highest posterior probabilities. Such graphs can be further compared with the neighborhood graph to see whether the spatial dependency patterns in observed data are actually consistent with the geographical neighborhoods.\\
\indent To gain further insight on the flexibility of our modeling approach, we examine the case in which a two-dimensional array $Y=\{ Y_{i_{1}i_{2}}\}$ is observed with the first dimension associated with $m_{1}$ diseases and the second dimension associated with $m_{2}$ areas. Under the framework of \cite{GeVo03,carlin-banerjee-2003} the matrix of counts $Y$ is modeled with a hierarchical Poisson model with random effects distributed as a multivariate CAR (MCAR) model \cite{Ma88}:
\begin{eqnarray}\label{eq:mcar}
 X^{\prime} \sim \normal_{m_{1}m_{2}}(0,[K_{1}\otimes (E_{W}-\rho W)]^{-1}).
\end{eqnarray}
This structure of the random effects assumes separability of the association structure among diseases from the spatial structure \cite{waller-carlin-2010}. The spatial autocorrelation parameter $\rho$ is the only parameter that controls the strength of spatial dependencies, while the precision matrix $K_{1}$ is not subject to any additional constraints on its elements. In our framework, the random effects $X^{\prime}$ follow a matrix-variate GGM prior obtained by taking $L=2$ in equation (\ref{eq:sepnorm}). The same separability of the association structure is assumed, but the precision matrices $K_{1}$ and $K_{2}$ follow G-Wishart hyper-priors as in equation (\ref{eq:sepnormprior}). The GGMs associated with the diseases are allowed to vary across all possible graphs with $m_{1}$ vertices, while the GGMs for the spatial structure can be modeled as we described earlier in this section.\\
\indent Until recently, the application of GGMs with a G-Wishart prior for the precision matrix in large scale Bayesian hierarchical models has been hindered by computational difficulties.  For decomposable graphs, the normalizing constant of the G-Wishart distribution is calculated with formulas \cite{roverato_2002,atay-kayis_massam_2005}, and a direct sampler from this distribution existed for several years \cite{carvalho_et_2007}. But similar results did not exist for non-decomposable graphs. Fortunately, new methodological developments give formulas for the calculation of the G-Wishart distribution for arbitrary graphs \cite{uhler-et-al-2014}, and also a direct sampler for arbitrary graphs \cite{lenkoski-2013}. With these key results, the MCMC sampler developed in Section \ref{sec:mcmc} can be significantly improved in its efficiency. The reversible jump algorithm that allows updates in the structure of the graphs associated the dimensions of a multi-way GGM can be subsequently refined to another transdimensional graph updating algorithm which bypasses the calculation of any normalizing constants of the G-Wishart distribution based on the double reversible jump algorithm of \cite{wang-li-2012,lenkoski-2013}. Moreover, the G-Wishart distribution can be replaced altogether in the specification of priors for Bayesian hierarchical spatial models with the graphical lasso prior of \cite{wang-2012}. The application of these new theoretical results to spatial health data is a very intense area of research.




\end{document}